\documentclass[aps,prl,twocolumn,showpacs,floatfix,preprintnumbers,nofootinbib,superscriptaddress]{revtex4-1}

\usepackage{graphicx}
\usepackage{color}
\usepackage{natbib}
\usepackage{multirow}

\usepackage{epsfig}
\usepackage{graphicx}
\usepackage{amsfonts}
\usepackage{amssymb}
\usepackage{latexsym}
\usepackage{amsmath}
\usepackage{hyperref}

\def\ee{\end{equation}}
\def\ba{\begin{eqnarray}}
\def\ea{\end{eqnarray}}

\def\bq{\begin{quote}}
\def\eq{\end{quote}}

 at 10truept

\newcommand{\beq}{\begin{equation}}
\newcommand{\eeq}{\end{equation}}
\newcommand{\beqa}{\begin{eqnarray}}
\newcommand{\eeqa}{\end{eqnarray}}
\newcommand{\bea}{\begin{eqnarray}}
\newcommand{\eea}{\end{eqnarray}}

\def\lesssim{~\mbox{\raisebox{-.6ex}{$\stackrel{<}{\sim}$}}~}

\def\ltap{\ \raise.3ex\hbox{$<$\kern-.75em\lower1ex\hbox{$\sim$}}\ }
\def\gtap{\ \raise.3ex\hbox{$>$\kern-.75em\lower1ex\hbox{$\sim$}}\ }
\def\gl{\ \raise.5ex\hbox{$>$}\kern-.8em\lower.5ex\hbox{$<$}\ }
\def\roughly#1{\raise.3ex\hbox{$#1$\kern-.75em\lower1ex\hbox{$\sim$}}}

\begin{document}

\title{New Early Dark Energy} 
\author{Florian Niedermann}
\email{niedermann@cp3.sdu.dk}
\author{Martin S. Sloth} 
\email{sloth@cp3.sdu.dk}
\affiliation{CP$^3$-Origins, Center for Cosmology and Particle Physics Phenomenology \\ University of Southern Denmark, Campusvej 55, 5230 Odense M, Denmark}


\begin{abstract}

New measurements of the expansion rate of the Universe have plunged the standard model of cosmology into a severe crisis. In this letter, we propose a simple resolution to the problem that relies on a first order phase transition in a dark sector in the early Universe, before recombination. This will lead to a short phase of a New Early Dark Energy (NEDE) component and can explain the observations. We model the false vacuum decay of the NEDE scalar field as a sudden transition from a cosmological constant source to a decaying fluid with constant equation of state. The corresponding fluid perturbations are covariantly matched to the adiabatic fluctuations of a sub-dominant scalar field that triggers the phase transition. Fitting our model to measurements of the cosmic microwave background (CMB), baryonic acoustic oscillations (BAO, and supernovae (SNe) yields a significant improvement of the best-fit compared with the standard cosmological model without NEDE. We find the mean value of the present Hubble parameter in the NEDE model to be $H_0=71.4 \pm 1.0 ~\textrm{km}\, \textrm{s}^{-1}\, \textrm{Mpc}^{-1}$ ($68\, \%$ C.L.). 

 \end{abstract}

\maketitle

\section{Introduction}

Recent measurements of the expansion of the Universe have led to an apparent crisis for the standard model of cosmology, the $\Lambda$CDM model. 
Within the $\Lambda$CDM model, we can calculate the evolution of the Universe from the earliest times until today, and until recently all our measurements were consistent with the model. In particular, we can use the measurements of the CMB radiation to infer the present value of the Hubble parameter $H_0$. If the $\Lambda$CDM model is correct, this value will have to agree with the value obtained by \emph{directly} measuring the expansion rate today using supernovae redshift measurements. Now, the problem is that the measurements, direct and indirect, do not agree, and this puts the $\Lambda$CDM model in a crisis.

The most precise {measurements} we have of the temperature fluctuations, polarization and lensing in the CMB radiation are from the Planck satellite, which, assuming the $\Lambda$CDM model, infers the value of the expansion rate today to be $H_0 = 67.36\pm 0.54~\textrm{km}\, \textrm{s}^{-1}\, \textrm{Mpc}^{-1}$ \cite{Aghanim:2018eyx}. Comparing that with the expansion rate measured from Cepheids-calibrated supernovae by the { SH${}_0$ES} team \cite{Riess:2019cxk}, $H_0 = 74.03\pm 1.42~\textrm{km}\, \textrm{s}^{-1}\, \textrm{Mpc}^{-1}$, there is a $4.4~\sigma$ discrepancy. Other measurements of the current Hubble rate, such as H${}_0$LiCoW \cite{Chen:2019ejq}, are also significantly discrepant with the Planck measurement~\cite{Verde:2019ivm}. 

The Planck measurement of the CMB is a very clean experiment with the systematics well under control, and it is therefore unlikely that there is non-understood systematics in the CMB measurements that can explain the discrepancy. The local supernova observations, on the other hand, involves astronomical distance measurements, which are notoriously difficult, and have been plagued by non-understood systematic errors in the past. Various possible sources of systematics have been considered extensively in the literature already~\cite{Davis:2019wet,Wojtak:2013gda,Odderskov:2017ivg,Wu:2017fpr}. 
So far, astronomers have no commonly accepted idea of possible systematic effects to explain the discrepancy, and an often echoed conclusion is that new physics beyond the $\Lambda$CDM model is required to resolve the tension (see e.g.,~\cite{Bernal:2016gxb,Knox:2019rjx}). While it is important to continue to look for possible systematic effects, in the present paper, we will rather consider a simple solution in terms of new physics. 

We will study the possibility that a \textit{first order phase transition} in a dark sector at zero temperature happened shortly before recombination in the early Universe.
Such a phase transition will have the effect of lowering an initially high value of the cosmological constant in the early Universe down to the value today, inferred from the measurement of $H_0$. Effectively this means that there has been an extra component of dark energy in the early Universe, providing a short burst of additional repulsion. Currently, an extra component of Early Dark Energy (EDE) seems to be a promising way to resolve the tension between the early and late measurements of $H_0$ \cite{Poulin:2018dzj,Poulin:2018cxd,Smith:2019ihp,Lin:2019qug,Kaloper:2019lpl,Alexander:2019rsc,Hardy:2019apu,Knox:2019rjx}. So far, people have typically considered a dynamical EDE component that disappears due to a second order phase transition of a slowly rolling scalar field.\footnote{The general idea of having an early dark energy component is older and dates back to \cite{Wetterich:2004pv,Doran:2006kp}.} Such scenarios have complications if monomial potentials are used both at background and perturbative level \cite{Agrawal:2019lmo,Smith:2019ihp}, as {one needs the potential to be steep and anharmonic at the bottom to end up with a sufficiently stiff fluid but also flat initially to achieve a sound speed $c_s^2 < 0.9$ for a large enough range of sub-horizon modes.} While this problem can be overcome by using specific terms from the non-perturbative form of the axion potential~\cite{Poulin:2018dzj,Poulin:2018cxd,Smith:2019ihp}, it represents a non-generic choice~\cite{Kaloper:2019lpl}.\footnote{For other  proposals to address the Hubble tension operational at late and/or early times see \cite{DiValentino:2016hlg,Kumar:2016zpg,Kumar:2017dnp,DiValentino:2017iww,DiValentino:2017zyq,Addison:2017fdm,Buen-Abad:2017gxg,Mortsell:2018mfj,Vagnozzi:2018jhn,Poulin:2018zxs,Yang:2018euj,DEramo:2018vss,Guo:2018ans,Aylor:2018drw,Kreisch:2019yzn,Raveri:2019mxg,Keeley:2019esp,DiValentino:2019exe,Gelmini:2019deq,Pan:2019gop, Vagnozzi:2019ezj,Visinelli:2019qqu,Pan:2019hac,DiValentino:2019ffd,Arendse:2019hev}.} 

On the other hand, we believe that a first order phase transition holds in it the potential to fully resolve the discrepancy between the early and late measurements of $H_0$ {much more naturally}. In addition, a first order phase transition will lead to different  experimental signatures in the details of the CMB and large-scale structure as well as gravitational waves.

Below we explore the simplest NEDE model. For more details and generalizations of the model, as well as a detailed comparison with other models, we refer the reader to our longer subsequent paper~\cite{long-one}.

\section{The Model}

In order to have a change in the vacuum energy due to a field that undergoes a first order phase transition, we will consider a scalar field with two non-degenerate minima at zero temperature. However, if the tunneling probability from the false to the true vacuum is initially high, the field will tunnel immediately and NEDE never makes a sizable contribution. On the other hand, once tunneling commences, we need a large rate in order to produce enough bubbles of true vacuum that will quickly collide. If the rate is too small, then part of the Universe will be in the true and part of it in the false vacuum, which will lead to large inhomogeneities ruled out by observations. We therefore require an additional sub-dominant trigger field that, at the right moment, makes the tunneling rate very high. Analogous to previously considered mechanisms for ending inflation in~\cite{Linde:1990gz,Adams:1990ds,Copeland:1994vg,Cortes:2009ej}, we will therefore consider models with a general potential of the form,
\bea\label{eq:action2}
V(\psi,\phi) &=&\frac{\lambda}{4}\psi^4+\frac{1}{2}\beta M^2 \psi^2\\ &&-\frac{1}{3}\alpha M \psi^3
+ \frac{1}{2}m^2\phi^2 +\frac{1}{2}\tilde\lambda \phi^2\psi^2  \,,
\nonumber
\eea
where $\psi$ is the tunneling field and $\phi$ is the trigger field. The {sub-dominant} trigger field will be frozen as long as its mass is smaller than the Hubble rate, but as soon as the Hubble rate drops below its mass, it will start decaying and this will trigger the tunneling of the $\psi$ field. For a second minimum to develop after the point of inflection, we need to impose $\alpha^2 > 4 \,\beta \lambda$, $\beta>0$. In Fig.~1, we show a 3D visualization of {the evolution of} the potential as the trigger field, $\phi$, starts evolving along the orange path opening up the new vacuum for $\psi$, to which it tunnels with high probability.


The decay rate per unit volume is $\Gamma = K \exp{\left( -S_E \right)}$, where $K$ is a determinant factor which is generically set by the energy scale of the phase transition~\cite{Callan:1977pt,Linde:1981zj} and $S_E$ is the Euclidian action corresponding to a so-called bounce solution~\cite{Coleman:1977py}. While it is possible to find an analytic expression in the thin wall limit {for a single field}, the general case requires a numerical approach. In \cite{long-one} we argue that a good approximation of the {Euclidian action (describing the potential as being effectively one-dimensional) can be written as 
\begin{align}\label{eq:SE}
S_{E} \approx \frac{4 \, \pi^2}{3 \lambda} \left( 2 - \delta_\text{eff}\right)^{-3} \left(\alpha_1 \delta_\text{eff} + \alpha_2 \delta_\text{eff}^2 + \alpha_3 \delta_\text{eff} ^3  \right)\,,
\end{align}  
with numerically determined coefficients~\cite{Adams:1993zs} $\alpha_1 = 13.832$, $\alpha_2 = -10.819$, $\alpha_3 = 2.0765$
and 
\begin{align}\label{eq:delta_phi}
\delta_\text{eff}(t) = 9\frac{ \lambda}{\alpha^2} \left( \beta + \tilde{\lambda} \frac{\phi^2(t)}{M^2}\right) \,.
\end{align}
We see that $S_{E}$ becomes large as $\delta_\text{eff} \to 2$ and vanishes as $\delta_\text{eff} \to 0$. As a result, the tunneling rate is suppressed when $\phi$ is frozen at a sufficiently large initial field value (corresponding to $\delta_\text{eff} > 9/4 \sim 2$) and becomes maximal as $\phi \to 0$ once the Hubble drag is released (corresponding to $\delta_\text{eff} \to 9\lambda \beta / \alpha^2 < 9/4$).

At early times, we require the transition rate to be highly suppressed, which fixes the initial value of the trigger field, $\phi_\text{ini}$, and can be satisfied consistently with the condition that $\phi_\text{ini}/M_{pl} \ll  1$, which is sufficient to ensure that the contribution of $\phi$ to the total energy density is sub-dominant.

\begin{figure}
\includegraphics[width=7cm]{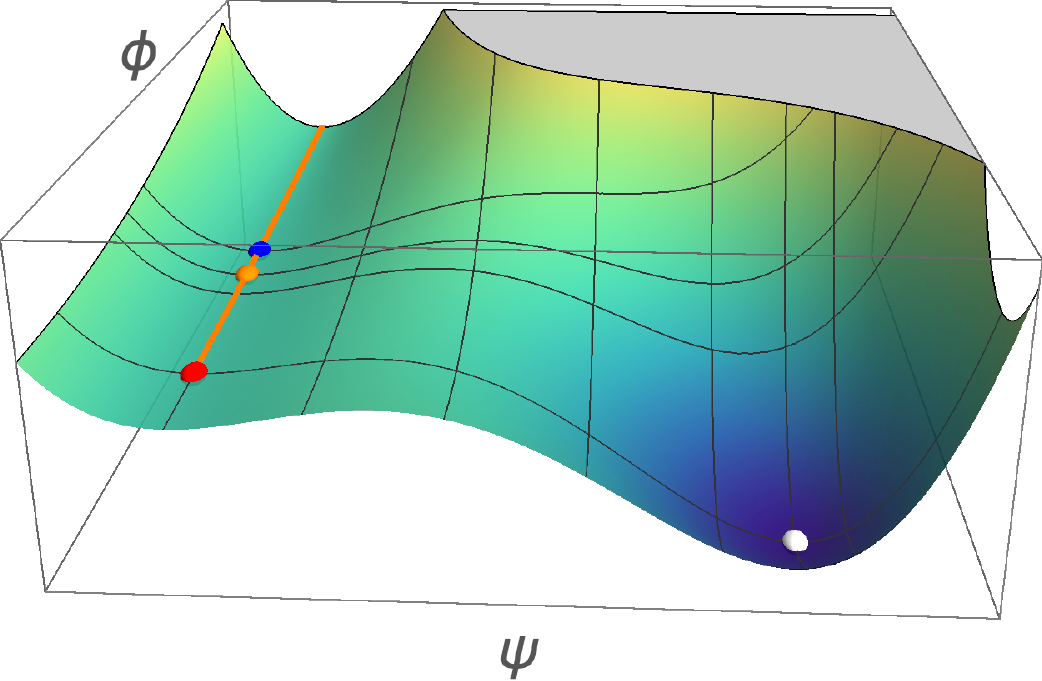}
\caption{Schematic plot of the two-field potential in \eqref{eq:action2}. For $H \lesssim m$, the field rolls along the orange line corresponding to $\psi = 0$. At the inflection point (blue dot) the potential (in $\psi$ direction) develops a second minimum which becomes degenerate shortly after (orange dot). The nucleation probability {increases towards} $\phi = 0$ (red dot). The true vacuum corresponds to the white dot. }
\label{fig:plot3d}
\end{figure}

Now, we also have to ensure that NEDE, given by the potential energy in the $\psi$ field, gives a sizable contribution to the energy budget at the time $t_*$ where bubble percolation of the $\psi$ vacuum becomes efficient. We can quantify it
in terms of the ratio $f_\text{NEDE} = \Delta V / \bar{\rho}(t_*)$, where $\Delta V$ is the liberated vacuum energy and $ \bar{\rho} $ the total energy density. 
 If the transition occurs at a redshift of order $z \sim 5000$, $\lambda \sim 0.1$, $\alpha \sim \beta \sim{\mathcal{O}(1)}$ and $f_\text{NEDE} \sim 0.1 $, we have $M \sim  \text{eV}$ and an ultra-light mass scale of order $m \sim 10^{-27} \text{eV}$. A microphysical model explaining the mass hierarchy between the $M$ and the $m$ scale would be a model of axion monodromy with two axion fields {(see \cite{Kaloper:2008fb} for a field theory version)}. Here, the masses are protected by softly broken shift symmetries.

We also have to make sure that the nucleation itself happens sufficiently quickly. To that end, we define the percolation parameter $p = \Gamma / H^4 \sim M^4 / m^4 \, {e^{-S_E}}$, where we approximated $K\sim M^4$. Provided $p  \gg 1$, a large number of bubbles is nucleated within one Hubble patch and one Hubble time.  In fact, for the above choice of parameters, the huge hierarchy between the scalar masses, $M^4/m^4 \sim 10^{108}$, implies that $p \gg 1$ only requires $S_{E} < 250$, which according to \eqref{eq:SE} and \eqref{eq:delta_phi} can be easily satisfied as $\phi \to 0$.
This means that percolation is extremely efficient and will cover the entire space with bubbles of true vacuum in a tiny fraction of a Hubble time. Therefore, we can treat it as an instantaneous process on cosmological time scales, which takes place at time~$t_*$.

As the space is being filled with bubbles of true vacuum, they expand and start to collide when they are of physical size  today {$\ll \text{Mpc}$}. Thus, they do not  induce anisotropies on scales large enough to be probed using CMB measurements. This phase is governed by complicated dynamics, which can be studied analytically only in simplified two-bubble scenarios as in \cite{Hawking:1982ga}. As part of the collision process, the complicated $\psi$ condensate starts to decay. Microscopically, the released free energy gets converted into anisotropic stress on small scales, which we expect, after partially being converted to gravitational radiation, to decay as $1/a^6$, similar to a stiff fluid component. We leave it for future work to substantiate this picture, which assumes a decoupling of small and large scales, through explicit numerical studies.

\section{Matching conditions}

We use a simple background model describing the instantaneous transition from a background fluid with an equation of state (e.o.s.)~parameter  that changes from $-1$ to $w_\text{NEDE}^*$, 

\begin{align}\label{eq:w}
w_\text{NEDE}(t) = 
\begin{cases}
-1   &\text{for} \quad t< t_* \,, \\ 
w_\text{NEDE}^*  &\text{for}\quad t>t_* \,,
\end{cases}
\end{align}
where the transition happens at time $t_*$. In terms of our field theory model in \eqref{eq:action2}, this corresponds to a situation where all of the liberated vacuum energy is transferred to a fluid with e.o.s.~parameter $w_\text{NEDE}^*$, and where according to the considerations above, we expect $1/3\leq w_\text{NEDE}^* \leq 1$. Describing the bubble wall condensate in terms of a fluid with a constant $w_\text{NEDE}^* $ is a simplifying assumption, which should ultimately be tested with lattice field theory techniques resolving the scalar field $\psi$ and its perturbations explicitly.\footnote{The importance of tracking field perturbations was highlighted recently in the case of the second-order, single-field EDE model in~\cite{Agrawal:2019lmo}, although after the NEDE transition, the frequency of the fluctuations is much higher, and the course grained fluid description is expected to be a better approximation.}

\subsection{Background Matching}

The above condition fixes the evolution of the background energy density uniquely, 
\begin{align} \label{eq:rho_EDE_bg}
\bar{\rho}_\text{NEDE}(t) = \bar{\rho}^*_\text{NEDE} \left( \frac{a(t_*)}{a(t)}\right)^{3[1+w_\text{NEDE}(t)]}\,,
\end{align}  
where $\rho_\text{NEDE}^* = f_\text{NEDE} \, \bar{\rho}_* = \text{const}$. The energy density of NEDE is normalized with respect to the true vacuum and continuous across the transition.
The discontinuity of a time dependent function $f(t)$ across the transition surface at time $t_*$ is denoted as

\begin{align}
\left[ f \right]_\pm = \lim_{\epsilon \to 0} \left[ f(t_*-\epsilon) - f(t_*+\epsilon) \right] \equiv f^{(-)} - f^{(+)}\;.
\end{align}
Applying this operation to the Friedmann equations, we then find
\begin{subequations}
\label{eq:matching_bg}
\begin{align}
\label{eq:JumpH}
\phantom{\left[ \dot H \right]_\pm} \left[ H \right]_\pm &= 0\;, \\
\left[ \dot H \right]_\pm &= 4 \pi G (1 + w^*_\text{NEDE}) \bar{\rho}_\text{NEDE}^* \, , 
\label{eq:JumpHdot}
\end{align}
\end{subequations}
where we used the continuity of the background energy density, $[\bar{\rho}]_\pm=0$,
which holds due to {\eqref{eq:rho_EDE_bg} and} the instantaneous character of the transition. The derivation of \eqref{eq:JumpHdot} also assumes that the e.o.s.~of all other fluid components (except for NEDE) is preserved during the transition.
Besides the NEDE component, we also {track the evolution of} the sub-dominant field  $\phi$  to turn on the phase transition.

\subsection{Perturbation Matching}

Before the decay we can set the perturbations of the NEDE fluid to zero as it behaves as a (non-fluctuating) cosmological constant. This raises the issue of how to initialize perturbations at time $t_*$. Moreover, since the transition is allowed to happen at a relatively late stage in the evolution of the primordial plasma (in the extreme case right before recombination), we cannot assume that all relevant modes are outside the horizon. 
In the specific case of our two-field model, we use the trigger field to define the transition surface  $\Sigma$, explicitly,  
\begin{align}
\phi(t_*, \mathbf{x})|_\Sigma = \text{const} \;.
\end{align}
This is motivated by the $\phi$ dependence of {the parameter} $\delta_\text{eff}$ in \eqref{eq:delta_phi} which controls the exponential in the tunneling rate through~\eqref{eq:SE}. 
As a consequence, fluctuations in $\phi$ lead to spatial variations of the time $t^*$ at which the decay takes place. These variations, $\delta \phi (t_*, \mathbf{x}) {= \phi(t_*, \mathbf{x})-\bar\phi(t_*)}$, then provide the initial conditions for the fluctuations in the NEDE fluid after the phase transition. 

In order to match the conventions used in the Boltzmann code community, we work in synchronous gauge,
\begin{align}\label{eq:sync}
ds^2 =  - dt^2  +a(t)^2 \left(\delta_{ij} + h_{ij} \right)dx^i dx^j \,,
\end{align}
where {in momentum space}
\begin{align}
h_{ij} = \frac{k_i k_j}{k^2} h + \left(\frac{k_i k_j}{k^2} - \frac{1}{3} \delta_{ij}\right)6\eta\,,
\end{align}
and $h=\delta^{ij} h_{ij}$. 
In the following we will make use of the equations for the metric perturbations that are first order in time derivatives~\cite{Ma:1995ey},
\begin{subequations}
\label{eq:Einstein}
\begin{align}
 \frac{1}{2} H \dot h - \frac{k^2}{a^2} \eta &=  4 \pi G \delta \rho \,,  \\
 \frac{k^2}{a^2} \dot \eta&= 4 \pi G \left(\bar{\rho} + \bar{p} \right) \frac{\theta}{a}\, ,
\end{align}
\end{subequations}
where $\left(\bar{\rho} + \bar{p} \right)\theta = \sum _i \left(\bar{\rho}_i + \bar{p}_i \right) \theta_i$ and $\delta \rho = \sum_i \delta \rho_i$ are the total divergence of the fluid velocity and the total energy density perturbation, respectively. The dynamical equations have to be supplemented with Israel's matching conditions~{\cite{Israel:1966rt,Deruelle:1995kd}}. They relate the time derivatives of $\eta$ and $h$ before and after the transition,

\begin{align}
\label{eq:matching}
\left[ \dot h \right]_\pm = -6 \left[ \dot \eta \right]_\pm = 6 \left[\dot H \right]_\pm \frac{\delta \phi(t_*, \mathbf{k})}{\dot{\bar{\phi}}(t_*)}\,,
\end{align}
where $\left[\dot H \right]_\pm$ is specified in \eqref{eq:JumpHdot}, and we used the residual gauge freedom in the synchronous gauge to bring the matching conditions on this simple form. We further find that all perturbations without a derivative, including the fluid sector, are continuous, i.e.\ $\left[ h \right]_\pm = \left[ \eta \right]_\pm =\left[ \delta_i \right]_\pm=\left[ \theta_i \right]_\pm  =  0 ${, where $\delta_i = \delta \rho_i / \bar{\rho}_i$}. 
This does not apply to NEDE perturbations because the derivation assumed that the e.o.s.~of a particular matter component $i$ is not changing during the transition, in contrast with \eqref{eq:w}. As argued before, we can consistently set 
\begin{align}
\delta_\text{NEDE} = \theta_\text{NEDE} = 0 \quad \text{for} \quad t < t_*\;.
\end{align}
We further introduce the notation $\delta_\text{NEDE}^{(+)} \equiv \delta_\text{NEDE}^{*}$ and  $\theta_\text{NEDE}^{(+)} \equiv \theta_\text{NEDE}^*$ to denote the fluctuations right after the transition. 
We can now evaluate the discontinuity of Einstein's equations \eqref{eq:Einstein} in order to fix $\delta_\text{NEDE}^*$ and $\theta_\text{NEDE}^*$, providing the initial conditions for the NEDE perturbations after the transition. Using \eqref{eq:matching} and \eqref{eq:JumpHdot}, we have
\begin{subequations}
\label{eq:EDE_ini}
\begin{align}
\delta^*_\text{NEDE} &= - 3 \left( 1 + w_\text{NEDE}^*\right) H(t_*) \frac{\delta \phi(t_*, \mathbf{k})}{\dot{\bar{\phi}}(t_*)} \; ,\\
\theta_\text{NEDE}^* &= \frac{k^2}{a(t_*)} \frac{\delta \phi(t_*, \mathbf{k})}{\dot{\bar{\phi}}(t_*)} \,.
\end{align}
\end{subequations}
These two equations together with the junction conditions \eqref{eq:matching} will allow us to consistently implement our model in a Boltzmann code.  In order to close the differential system of the perturbed fluid equations, we set the rest-frame sound speed~\cite{Hu:1998kj} in the NEDE fluid to $c_s^2 = w_\text{NEDE}^*$. 

\section{Data Analysis and Results}

In order to fit the NEDE model to the CMB data, we have incorporated it into the Boltzmann code\footnote{The adapted {\tt CLASS} code is publicly available on GitHub:  \url{https://github.com/flo1984/TriggerCLASS} } {\tt CLASS} \cite{Lesgourgues:2011re,Blas:2011rf}.  To that end, we made the simplifying assumption that all liberated vacuum energy is ultimately converted to small scale anisotropic stress and gravitational radiation described as a fluid with \ $1/3\leq w^*_\text{NEDE} \leq 1$. As a specific choice for our data fit, we take the midpoint $w^*_\text{NEDE} = 2/3 (= c_s^2)$, which we relax in our subsequent paper~\cite{long-one}. In accordance with our microscopic model the decay is triggered shortly before $\phi=0$, where for definiteness we take $H/m = 0.2$ (which avoids a tuning and is still compatible with a quick decay). The sub-dominant trigger field and its perturbations are evolved explicitly and matched to the fluid perturbations through~\eqref{eq:EDE_ini}. This scenario also assumes that there are no sizeable oscillations in $\phi$ around the true vacuum, which could give rise to an additional sub-dominant dust component. A more detailed discussion of the corresponding microscopic constraints is provided in Sec.~IID~in~\cite{long-one}.

The cosmological parameters are then extracted with the Monte Carlo Markov chain code {\tt MontePython}~\cite{Audren:2012wb,Brinckmann:2018cvx}{, employing a Metropolis-Hastings algorithm}. We perform a model comparison by computing the difference in Bayesian evidence $\Delta \log B = \log B(\text{NEDE}) - \log B$($\Lambda$CDM) using the {\tt MultiNest} algorithm (evidence tolerance 0.1 and 1000 live-points)~\cite{Feroz:2008xx,Feroz:2013hea,Buchner:2014nha}. Compared to $\Lambda$CDM, we introduce {two} new parameters: the fraction of NEDE before the decay,  $f_\text{NEDE} = \bar{\rho}_\text{NEDE}^* / \bar{\rho}(t_*)$, and the logarithm of the mass of the trigger field $\log_{10} (m \times \text{Mpc})$, which defines the redshift at decay time, $z_*$, via $H(z_*) = 0.2\, m$. In total, we vary eight parameters $\{\omega_{b },\, \omega_\text{cdm},\, h,\, \ln10^{10}A_{s },\, n_{s },\, \tau_\text{reio },\, f_\text{NEDE}, \, \log_{10} (m \times \text{Mpc}) \}$, on which we impose flat priors. The neutrino sector is modeled in terms of two massless  and one massive species with $M_\nu = 0.06 \, \text{eV}$. We impose the initial value $\phi_\text{ini} / M_{pl} = 10^{-4} $ to make sure that the trigger field is always sub-dominant and the tunneling rate at early times sufficiently suppressed.  
 
 We will use the following data sets: 
 the most recent SH$_0$ES measurement, which is $H_0 = 74.03 \pm 1.42 \, {\text{km}\,  \text{s}^{-1} \text{Mpc}^{-1}}$~\cite{Riess:2019cxk}; the Pantheon data set \cite{Scolnic:2017caz} comprised of 1048 SNe Ia in a range $0.01 < z < 2.3$; the large-$z$ BOSS DR 12 anisotropic BAO {and growth function} measurements at redshift $z = 0.38$, $0.51$ and $0.61$ based on the CMASS and LOWZ galaxy samples~\cite{Alam:2016hwk}, as well as small-z, isotropic BAO measurements of the 6dF Galaxy Survey \cite{Beutler:2011hx} and the SDSS DR7}main Galaxy sample \cite{Ross:2014qpa} at $z=0.106$ and $z=0.15$, respectively (collectively referred to as BAO); the Planck 2018 TT, TE, EE and lensing likelihood \cite{Aghanim:2019ame} with all nuisance parameters; constraints on the primordial helium abundance from \cite{Aver:2015iza} (referred to as BBN). We perform one likelihood analysis with all data sets combined (see red contours in Fig.~\ref{fig:f_NEDE}), one where we only exclude the SH$_0$ES value (turquoise contours) and one with Planck (TT, TE, EE and lensing) alone (orange contours). For the latter two, we fix $\log_{10}(m \times \text{Mpc}) = 2.58$ in order to avoid sampling volume artifacts in the $f_\text{NEDE} \to 0$ limit. While we provide an exhaustive discussion of this issue and also results without fixing $\log_{10}(m)$ in our subsequent paper~\cite{long-one}, here, we highlight the main findings~\cite{supp_material}.

{For the analysis with all data sets, the best-fit improves by $\Delta \chi^2= -15.6$ compared to $\Lambda$CDM. This improvement is shared between SH$_0$ES [$\Delta\chi$(SH$_0$ES) =  -13.8] and the other data sets  [$\Delta\chi$(w/o SH$_0$ES) =  -1.8]. This observation is crucial at it shows that NEDE does not lead to new tensions. Instead, it also improves the overall fit to the other data sets.\footnote{There is a slight degradation of the large-$z$ BAO data set of $\Delta \chi^2$(large-$z$ BAO + LSS) = 0.9, which we will explore in our future work about large-scale structure within NEDE~\cite{Niedermann:2020qbw}.}
Moreover, we find $H_0 = 71.4 \pm 1.0~\textrm{km}\, \textrm{s}^{-1}\, \textrm{Mpc}^{-1}$.  The decay takes place at $z_* =4920_{-730}^{+620} $, and  there is  a non-vanishing NEDE fraction $f_\text{NEDE}= 12.6_{-2.9}^{+3.2}\, \%$, excluding $f_\text{NEDE}=0$ with a $4.3\, \sigma$ significance.  This is also supported by the Bayesian evidence measure, which amounts to $\Delta B = 5.5$, corresponding to a ``very strong'' evidence on Jeffreys' scale~\cite{Efron}.}

This picture is further solidified by our runs without SH$_0$ES, which lead to a (very similar) fit improvement of  $\Delta \chi^2= -3.1$ (Planck) and  $\Delta \chi^2= -2.9$  (Planck+BAO+Pantheon+BBN). In both cases, we find a $1.9\, \sigma$ evidence for a non-vanishing value of $f_\text{NEDE}$ and a ``weak'' (but positive) Bayesian evidence of $0.6< \Delta B < 0.7$ [when fixing $\log_{10}(m)$].  The mean values are $H_0 = 69.5^{+1.1}_{-1.5}~\textrm{km}\, \textrm{s}^{-1}\, \textrm{Mpc}^{-1}$  and $H_0 = 69.6^{+1.0}_{-1.3}~\textrm{km}\, \textrm{s}^{-1}\, \textrm{Mpc}^{-1}$, respectively, which brings the Hubble tension down to $2.5\, \sigma$, in turn justifying our joint analysis. In short, NEDE introduces an approximate degeneracy in the $f_\text{NEDE}$ vs $H_0$ plane (see Fig.~\ref{fig:f_NEDE}). The SH$_0$ES measurement is then needed to select NEDE as the favored model.

\begin{figure}
\includegraphics[width=8cm]{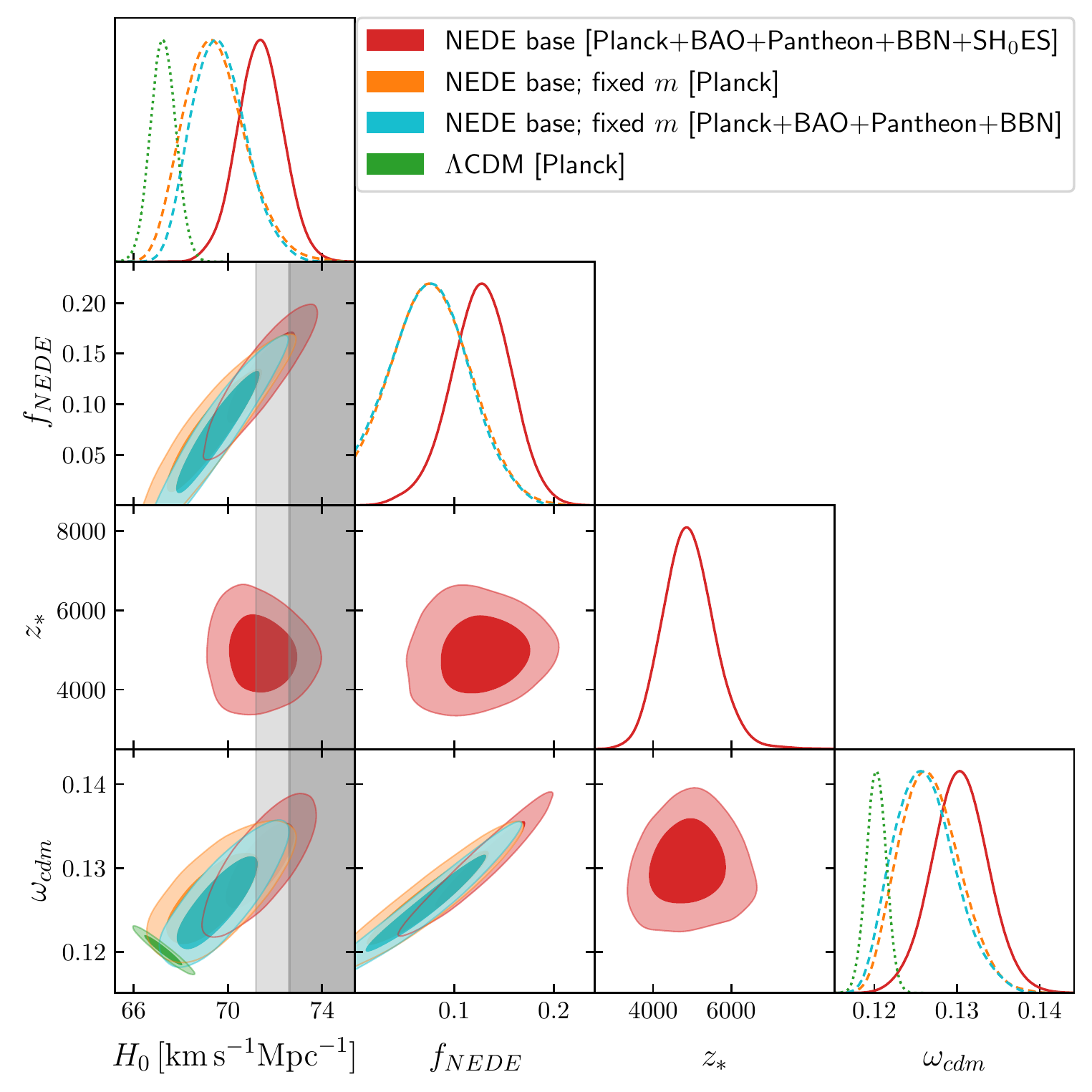}
\caption{Covariances and posteriors of $H_0$, $f_\text{NEDE}$, $z_*$ and $\omega_\text{cdm}$ for our combined analyses. The $68 \%$ and $95 \%$ C.L.\  correspond respectively to the light and dark shaded regions. The SH$_0$ES value is represented as the vertical gray bands.}
\label{fig:f_NEDE}
\end{figure}

\section{Conclusions}

We have studied a NEDE model where the decay of NEDE happens through a first order phase transition. This makes our model unique compared to older EDE models (which all rely on a second order phase transition) both from a theoretical and phenomenological perspective. The NEDE model holds in it the potential to fully resolve the discrepancy in $H_0$ as inferred from early CMB and BAO measurements and late time {distance ladder} measurements. Our first most simplified implementation of the model (fixing as many free parameters as possible by making simple assumptions) already yields a significant improvement in the fit over the $\Lambda$CDM model of {$\Delta \chi^2 = -15.6$} {when including the SH$_0$ES measurement of $H_0$. Crucially, this does not compromise the fit to the other data sets. Correspondingly, without including the SH$_0$ES prior on $H_0$, the Hubble tension is reduced to the $2.5 \, \sigma$  level}. 
We expect that the model will fit the data even better when the simplifying assumptions made in the present short paper are dropped in future work.

\begin{acknowledgments}
We thank {Edmund Copeland, Jos\'e Espinosa, Nemanja Kaloper, Antonio Padilla and Kari Rummukainen} for discussions {and/or} useful comments on the manuscript.
This work is supported by Villum Fonden grant 13384. CP3-Origins is partially funded by the Danish National Research Foundation, grant number DNRF90.
\end{acknowledgments}

\end{document}